\documentstyle[fleqn,espcrc2,epsf]{article}

\def\beq{\begin{equation}}
\def\eeq{\end{equation}}
\def\bea{\begin{eqnarray}}
\def\eea{\end{eqnarray}}

\def\myfigure#1#2#3#4#5#6#7{
  \begin{figure}[hbtp]
  \hskip #1
  \vskip #2
    \begin{center}
      \epsfxsize=#3
      \leavevmode
      \epsffile{#4}
    \end{center}
    \vskip #5
    \caption{#6}
    \label{#7}
  \end{figure} }

\newcommand{\mb}{\overline{m}_b}

\begin{document}

%%%%%%%%% cover page %%%%%%%%%
\begin{titlepage}
\renewcommand{\thefootnote}{\fnsymbol{footnote}}
\begin{flushright}
   FTUV/98-77 \\ IFIC/98-78
\end{flushright}
\par \vspace{10mm}
\begin{center}
{\Large \bf
Heavy quark mass effects in $e^+e^-$ into three jets
\footnote{to be published in the Proceedings of the High
Energy Physics International Euroconference on Quantum
Chromodinamics ({\it QCD '98}),
Montpellier, France, 3-9 Jul 1998.
Ed. S. Narison, Nucl Phys. B (Proc. Suppl.).}}
\end{center}
\par \vspace{2mm}
\begin{center}
Mikhail Bilenky$^a$\footnote{On leave from JINR, 141980 
Dubna, Russian Federation}
Germ\'an Rodrigo$^b$\footnote{On leave from Departament de
F\'{\i}sica Te\`orica, IFIC,
CSIC-Universitat de Val\`encia, 46100 Burjassot, Val\`encia, Spain},
Arcadi Santamaria$^c$ and
\vspace*{0.5cm} \\
$^a$~Institute of Physics, AS CR, 18040 Prague and \\
Nuclear Physics Institute, AS CR, 25068 \v{R}e\v{z}(Prague), Czech Republic \\
$^b$~INFN-Sezione di Firenze, Largo E. Fermi 2, 50125 Firenze, 
Italy \\
$^c$~Departament de F\'{\i}sica Te\`orica, IFIC,
CSIC-Universitat de Val\`encia, 46100 Burjassot, Val\`encia, Spain
\end{center}
\par \vspace{2mm}
\begin{center} {\large \bf Abstract} \end{center}
\begin{quote}
Next-to-leading order calculation for
three jet heavy quark production in $e^+e^-$ collisions,
including complete quark mass effects, 
is reviewed. Its applications at LEP/SLC are also discussed. 
\end{quote}
\vspace*{\fill}
%\begin{flushleft}
%     \today
%\end{flushleft}
\end{titlepage}
%%%%%%%%%%%%%%%%%%%%%

\newpage\addtocounter{footnote}{-3}
\pagestyle{empty}

% declarations for front matter
\title{Heavy quark mass effects in $e^+e^-$ into three jets
%\thanks{
%}
}
\author{
Mikhail Bilenky~\address{
Institute of Physics, AS CR, 18040 Prague and \\ 
Nuclear Physics Institute, AS CR, 25068 \v{R}e\v{z}(Prague),
Czech Republic}\thanks{On leave from JINR, 141980 
Dubna, Russian Federation.}, 
Germ\'an Rodrigo~\address{
INFN-Sezione di Firenze, Largo E. Fermi 2, 50125 Firenze, 
Italy}\thanks{On leave from Departament de F\'{\i}sica Te\`orica, IFIC,
CSIC-Universitat de Val\`encia, 46100 Burjassot, Val\`encia, Spain.} and
Arcadi Santamaria~\address{
Departament de F\'{\i}sica Te\`orica, IFIC,
CSIC-Universitat de Val\`encia, 46100 Burjassot, Val\`encia, Spain} 
}

\begin{abstract}
Next-to-leading order calculation for
three-jet heavy quark production in $e^+e^-$-collisions,
including complete quark mass effects,
is reviewed. Its applications at LEP/SLC are also discussed. 
\end{abstract}

\maketitle

%\section{Introduction}

The importance of the corrections due to the mass of the heavy quark in the 
jet-production in $e^+e^-$-collisions has been already seen 
in the early tests of the
flavour independence of the strong coupling constant \cite{delphi,lep}. The 
final high  
precision of the LEP/SLC experiments required accurate account for the
bottom-quark mass in the theoretical 
predictions. 
If quark mass effects are neglected,
the ratio $\alpha_s^b / \alpha_s^{uds}$ measured from the analysis of different
three-jet event-shape observables is shifted away from unity
up to $8\%$ \cite{opal98} (see also \cite{delphi97}).

Sensitivity of the three-jet observables to the value of the 
heavy quark mass allowed to consider the possibility \cite{juan,brs95} 
of the determination of the b-quark mass from LEP data, 
assuming the universality of $\alpha_s$.
In a recent analysis of three-jet events,
DELPHI measured the mass of the b-quark, $m_b$, 
for the first time far above the production threshold \cite{delphi97}. 
This result is in a 
good agreement
with low energy determinations of $m_b$
using QCD sum rules and lattice QCD from 
$\Upsilon$ and $B$-mesons spectra (for recent results see e.g. \cite{lowenergy}) 
The agreement between high and low energy
determinations of the quark mass is rather impressive 
as non-perturbative parts are very different in the two cases.

In this contribution we will discuss some aspects of the next-to-leading order
(NLO) calculation of the decay $Z \rightarrow 3 jets$ with massive quarks,
necessary for the measurements of the bottom-quark mass at the $Z$-peak.
Recently such calculations were performed independently by three 
groups \cite{rsb97,bbu97,no97}.

The first question we would like to answer whether it is not at all
surprising that LEP/SLC observables are sensitive to 
$m_b$ as the main scale involved is
the mass of the $Z$-boson, $M_Z \gg m_b$.
Indeed, the quark-mass effects for an inclusive observable such as the total
width $Z \rightarrow b \bar{b}$ are negligible. Due to 
Kinoshita-Lee-Nauenberg theorem 
such observable does not contain mass 
singularities and a quark-mass appears in the ratio 
$\overline{m}_q^2(M_Z)/M_Z^2 \approx 10^{-3}$, where using $\overline{MS}$
running mass at the $M_Z$-scale takes into account the bulk of the NLO 
QCD corrections \cite{dkz90,brs95}.

However, the situation with more exclusive observables is different.
Let's consider the simplest process, $Z \rightarrow \overline{b}b g$,
which contributes to three-jet final state at the leading
order (LO). When the energy of the radiated gluon approaches zero, the  
process has an infrared (IR) divergence and in order to have an
IR-finite prediction, some kinematical restriction should be introduced
in the phase-space integration to cut out the troublesome region.
In $e^+e^-$-annihilation that is usually done by applying the so-called
jet-clustering algorithm with a jet-resolution parameter, $y_c$ (see
\cite{mls98} for recent discussion of jet-algorithms in $e^+e^-$). 
Then the transition probability in the three-jet 
part of the phase-space will have contributions as large as 
$1/y_c \cdot (m_b^2/M_Z^2)$,
where $y_c$ can be rather small, in the range $10^{-2} - 10^{-3}$.
Then one can expect a significant enhancement of 
the quark-mass effects, which can reach several percents.

The convenient observable for studies of the mass effects in the three-jet
final state, proposed some time ago \cite{delphi,brs95}, is defined as follows
\bea
\label{r3bd}
& & R_3^{bd}=\frac{\Gamma^b_{3j}(y_c)/\Gamma^b}{\Gamma^d_{3j}(y_c)/\Gamma^d}
\\ \nonumber
& &=1+r_b \left( b_0(y_c,r_b) + \frac{\alpha_s}{\pi} b_1(y_c,r_b  
) \right)
\eea
where $\Gamma^q_{3j}$ and $\Gamma^q$ are three-jet and total decay
widths of the $Z$-boson into quark pair of  flavour $q$,
$r_b=m_b^2/M_Z^2$. Note that above expression is not an expansion
in $r_b$.

The LO function, $b_0$, is plotted in fig.~1  
for four different jet algorithms. 
\vskip -2.cm
%%%%%%%%%%%%%%%
\myfigure{0.cm}{0.cm}{8.2cm}{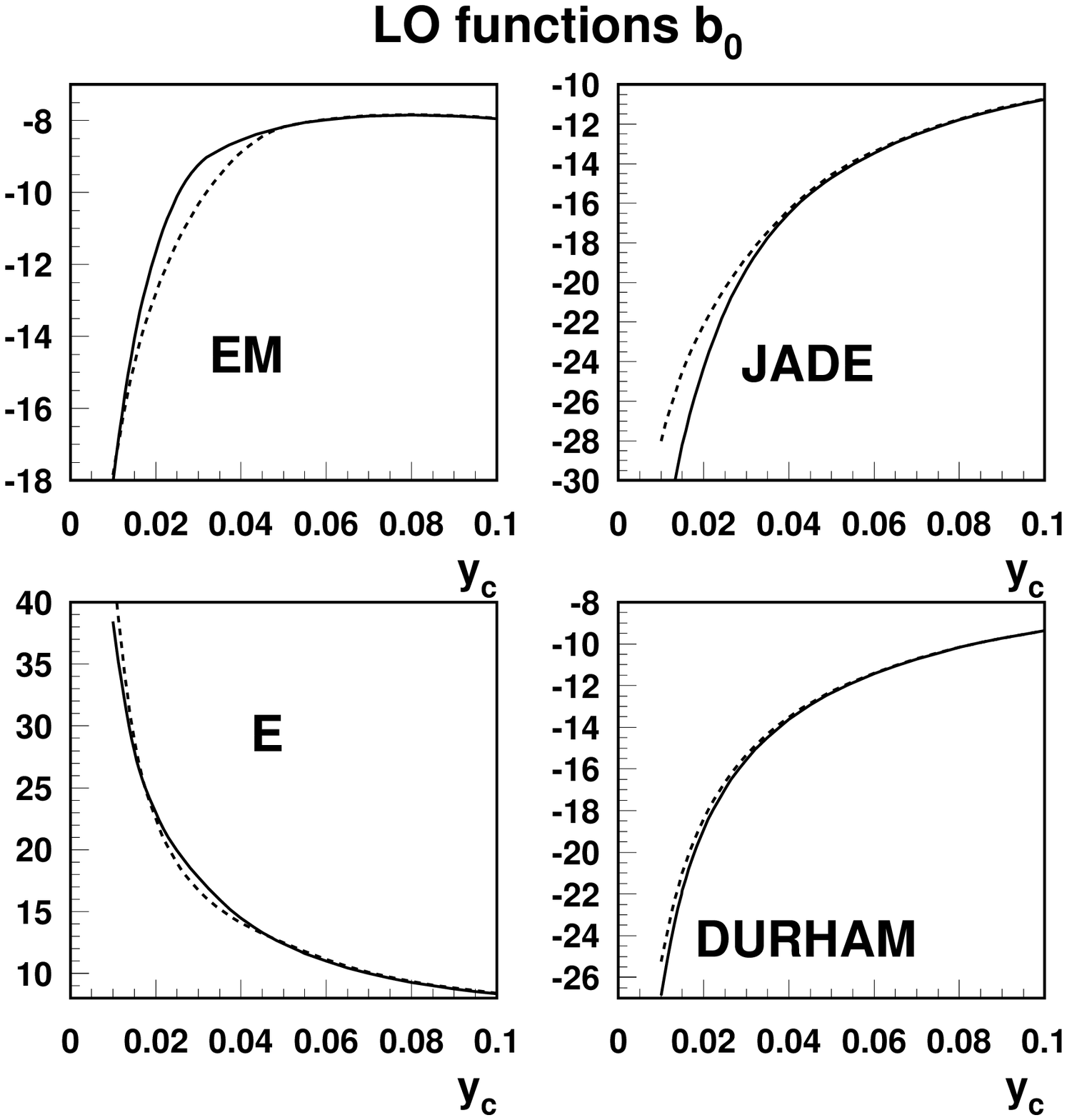}{-1.cm}
{\it LO contribution to the ratio $R_3^{bd}$
as a function of $y_c$ (see eq.(\ref{r3bd}) for the definition)
for $m_b=3 GeV$ (dashed curve) and $m_b=5 GeV$ (solid curve).}{fig:b0}
%%%%%%%%%%%%%%%
\vskip -8mm
Together with well known JADE, E and DURHAM schemes
we consider the so-called EM algorithm \cite{brs95} with
a resolution parameter $y_{ij}=2p_i p_j/s$ and which was 
used for analytical calculations in the massive case \cite{brs95}.
The main observation from fig.~1 is that for $ y_c>0.05$,
$b_0$ is almost independent of the value of $m_b$ for all schemes. 
Although, this remains true
also for smaller $y_c$ in DURHAM and E schemes, there is a noticeable
mass dependence in JADE and EM schemes.

Note that $b_0$ is positive for E-scheme. That contradicts
the intuitive expectations that a heavy quark should radiate less than a light one.
This unusual behavior is due to the definition
of the resolution parameter in E-scheme, $y_{ij}=(p_i+p_j)^2/s$, which has
significantly different values for partons with the same momenta 
in the massive and massless cases, and it can 
be used as a consistency check of the data.

In what follows we restrict ourselves to DURHAM scheme,
the one used in the experimental analysis \cite{delphi97},
and $b_0$ can be interpolated as:
$b_0=b_0^{(0)}+b_0^{(1)} \ln y_c+b_0^{(2)} \ln^2 y_c$.
In the LO calculations we can not specify what value
of the b-quark mass should be taken in the calculations: all quark-mass
definitions
are equivalent (the difference is due to the higher orders in $\alpha_s$). 
One can use, for example, the pole mass $M_b \approx 4.6 GeV$
or the $\overline{MS}$-running mass $\overline{m}_b(\mu)$ at any scale 
relevant to the problem, $m_b \le \mu \le M_Z$, with $\overline{m}_b(m_b)
\approx 4.13 GeV$ and $\overline{m}_b(M_Z) \approx 2.83 GeV$.
As a result, the spread in LO predictions
for different values of b-quark mass is significant, 
the LO prediction is not accurate enough and
the NLO calculation should be done.

At the NLO there are two different contributions: from one-loop  
corrections to the three-parton decay, $Z \rightarrow b \overline{b} g$
and tree-level four-parton decay, $Z \rightarrow
b \overline{b} g g$ and 
$ Z \rightarrow b \overline{b} q \overline{q},~q=u,d,s,c,b$
integrated over the three-jet region of the four-parton phase-space.
In the NLO calculation one has to deal with divergences,
both ultraviolet \footnote{The ultraviolet divergences in 
the loop-contribution 
are cancelled after the renormalization of the
parameters of the QCD Lagrangian.}
and infrared  which appear at the intermediate
stages. 
The sum of the one-loop and tree-level contributions is, however, IR finite.
We would like to stress that the structure of the NLO corrections in the
massive case is completely different from the ones in the massless
case \cite{ert81}. That is due to the fact
that in the massive case, part of the collinear divergences, those
associated with the gluon radiation from the quarks, are softened
into $\ln r_b$ and only collinear 
divergences associated with gluon-gluon splitting remain.

In the NLO calculations one should specify the quark mass definition.
It turned out technically simpler to use a mixed renormalization
scheme which uses on-shell definition for the quark mass 
and $\overline{MS}$ definition for the strong coupling. 
Therefore, physical quantities
are originally expressed in terms of the pole mass. 
It can be perfectly used in perturbation theory, 
however, in contrast to the pole
mass in QED, the quark pole mass is not a physical parameter. The
non-perturbative corrections to the quark self-energy bring an
ambiguity of order $\approx 300 MeV$ (hadron size) to the physical position 
of the pole of the quark propagator. 
Above the quark production 
threshold, it is natural to use the running mass definition
(we use $\overline{MS}$). The advantage of this definition is
that $\overline{m}_b(\mu)$ can be used for $\mu \gg m_b$.
\vskip -2.cm
%%%%%%%%%%%%%%%
\myfigure{0.cm}{0.cm}{6cm}{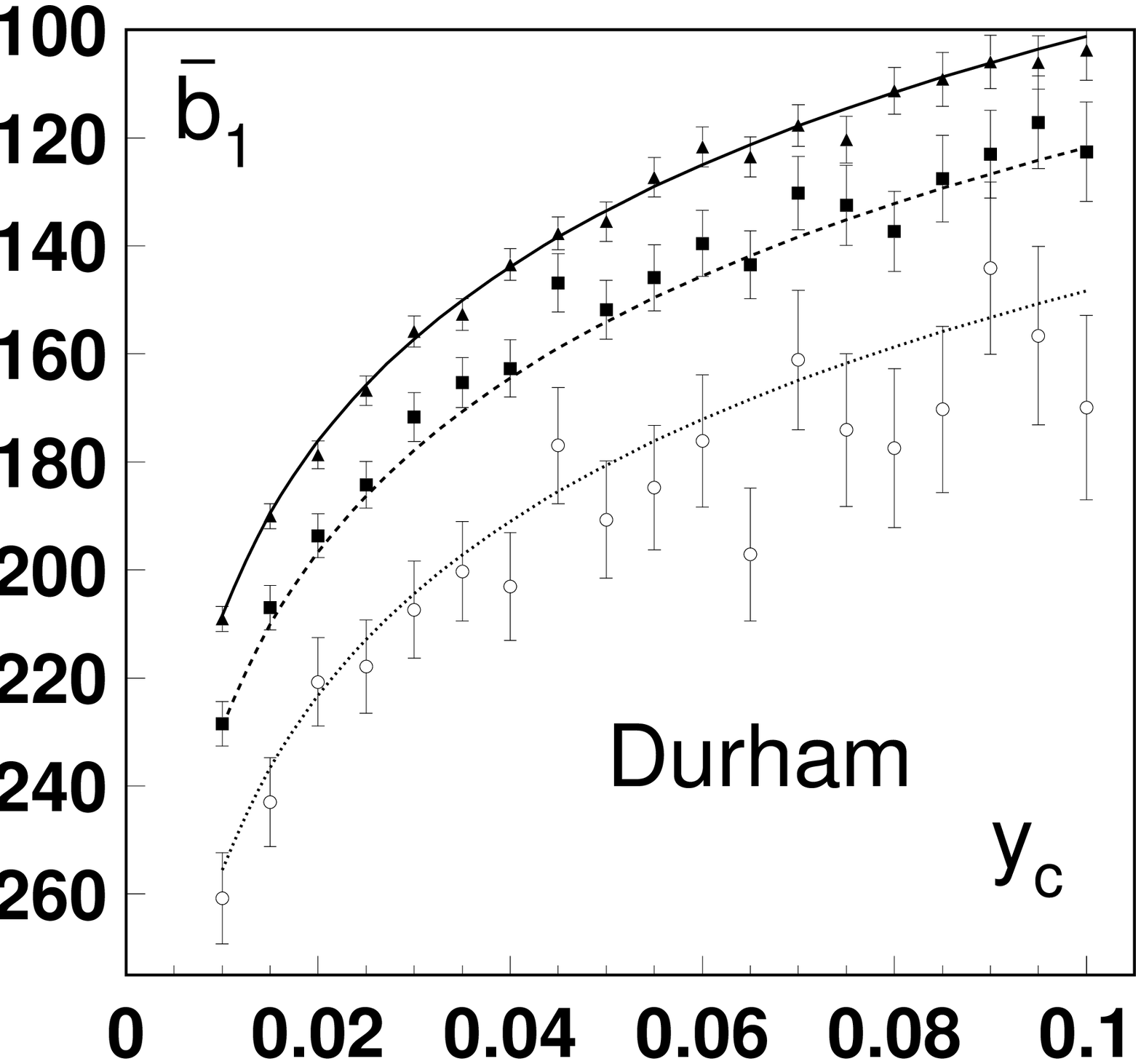}{-1.cm}
{\it NLO function $\overline{b}_1$ for different $m_b$
(see eqs.(\ref{r3bd}),(\ref{r3bdrun}) for the definition and text for details).
The errors are due to numerical
integrations.}{fig:b1}
%%%%%%%%%%%%%%%
\vskip -8mm
The pole, $M_b$, and the running masses of the quark
are perturbatively related
\beq
M_b=\overline{m}_b(\mu)\left[1+\frac{\alpha_s}{\pi}\left(\frac{4}{3}
-\ln\frac{m_b^2}{\mu^2}\right)\right]~.
\label{pole-run}
\eeq
We use this one-loop relation to pass from the pole mass to the running one,
which is consistent with our NLO calculations. To match needed precision 
we have to use this equation for values of $\mu$ about $m_b$.
Then we can use one-loop renormalization group improved equation
in order to define the quark mass at the higher scales.
Substituting eq.(\ref{pole-run}) into definition eq.(\ref{r3bd}) we have
\begin{eqnarray}
& &R_3^{bd}(y_c,\mb(\mu),\mu)=
\nonumber \\
& &1+\overline{r}_b(\mu)\left[b_0 
+ \frac{\alpha_s(\mu)}{\pi} \left(\overline{b}_1
-2b_0\ln\frac{M_Z^2}{\mu^2}\right)\right]
\label{r3bdrun}
\end{eqnarray}
with $\overline{b}_1=b_1+b_0(8/3-2\ln r_b)$ and $\overline{r}_b=\overline{m}_b^2/M_Z^2$.

In fig.~2 we show the NLO function $\overline{b}_1(y_c,r_b)$ calculated for
three different values of the quark mass : $3GeV$ (open circles), $4GeV$ (squares)
and $5GeV$ (triangles).
\vskip -2.cm
%%%%%%%%%%%%%%%
\myfigure{0.in}{-0.5cm}{6cm}{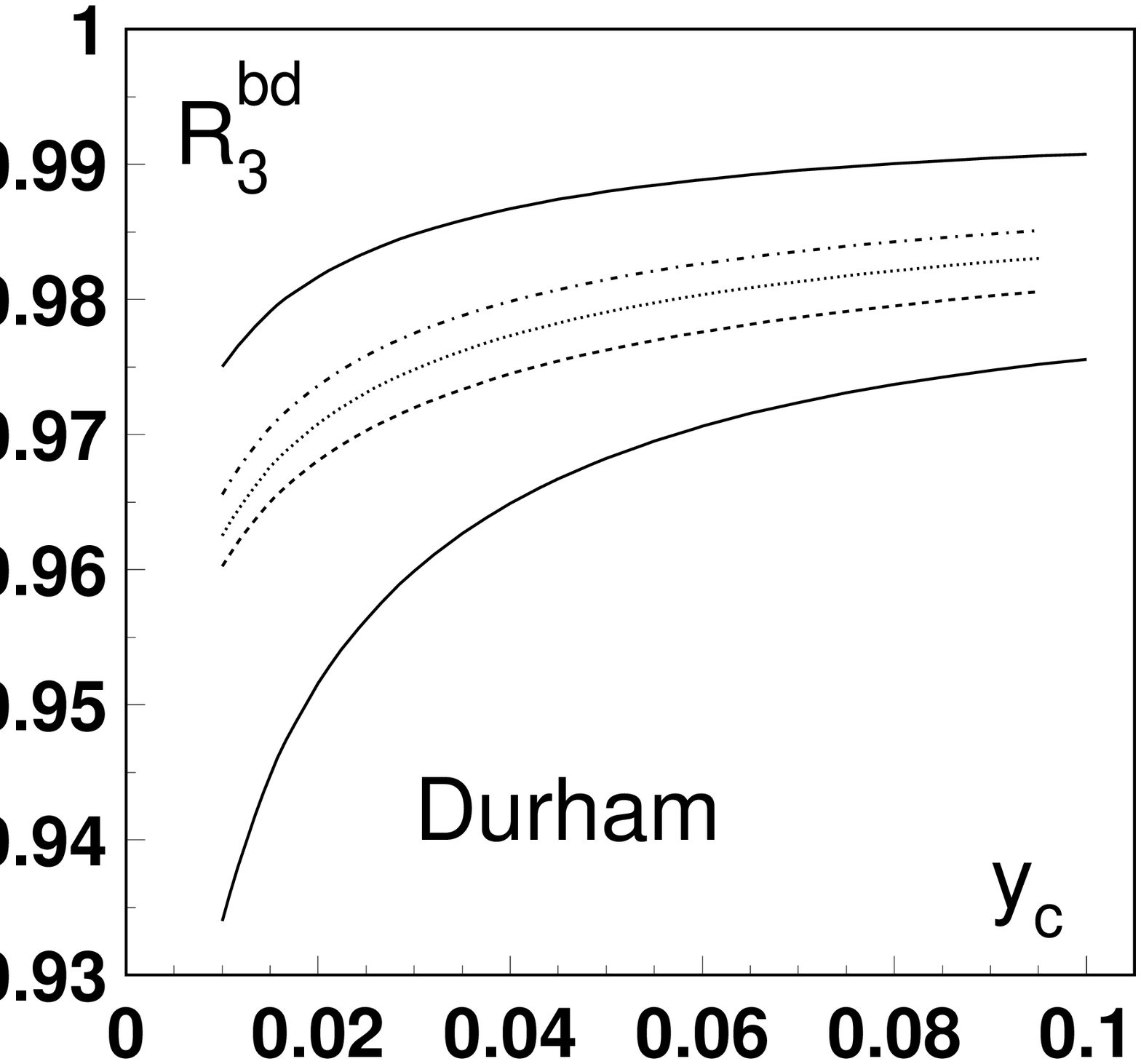}{-1.cm}
{\it The ratio $R_3^{bd}$ (eq. (\ref{r3bd})). Solid curves - LO predictions,
dashed curves give the NLO results (see text for details).}
{fig:r3bd}
%%%%%%%%%%%%%%%
\vskip -8mm
In contrast to the $b_0$,
one sees a significant residual mass
dependence in $\overline{b}_1$, which can not be neglected. The solid lines
in fig.~2
represent a fit by the function: $\overline{b}_1=\overline{b}_1^{(0)}
              +\overline{b}_1^{(1)} \ln y_c 
              +\overline{b}_1^{(2)} \ln r_b $
performed in the range $ 0.01 \le y_c \le 0.1$ 
The quality of this interpolation is very good and
the main residual $m_b$ dependence in $\overline{b}_1$ is taken into account
by $\ln r_b$ term. Inclusion of higher powers of $\ln r_b$ does
not improve the fit.

Fig.~3 presents theoretical predictions 
in the DURHAM scheme
for the $R_3^{bd}$ observable measured by DELPHI\cite{delphi97}.
The solid lines are LO predictions for the b-quark mass, $m_b=\mb(M_Z)=2.83 GeV$ 
(upper curve) and $m_b=M_b=4.6 GeV$ (lower curve). The dashed curves
give NLO results for different values of scale $\mu:~10,~30,~91 GeV$.
One sees that NLO curve for large scale is naturally closer to LO curve
for $\mb(M_Z)$ and for smaller scale is closer to the LO one with
$m_b=M_b$.
 
Fig.~4 illustrates the scale dependence of $R_3^{bd}$ for $y_c=0.02$. 
By studying the scale dependence, which is a reflection
of the fixed order calculation, we can estimate the uncertainty
of the predictions. The dashed-dotted curve gives $\mu$-dependence
when eq. (\ref{r3bd}) was used, so it is $\mu$-dependence due to
renormalization of
the strong coupling constant, $\alpha_s$. 
\vskip -2.cm
%%%%%%%%%%%%%%%
\myfigure{-1.in}{-0.5cm}{6cm}{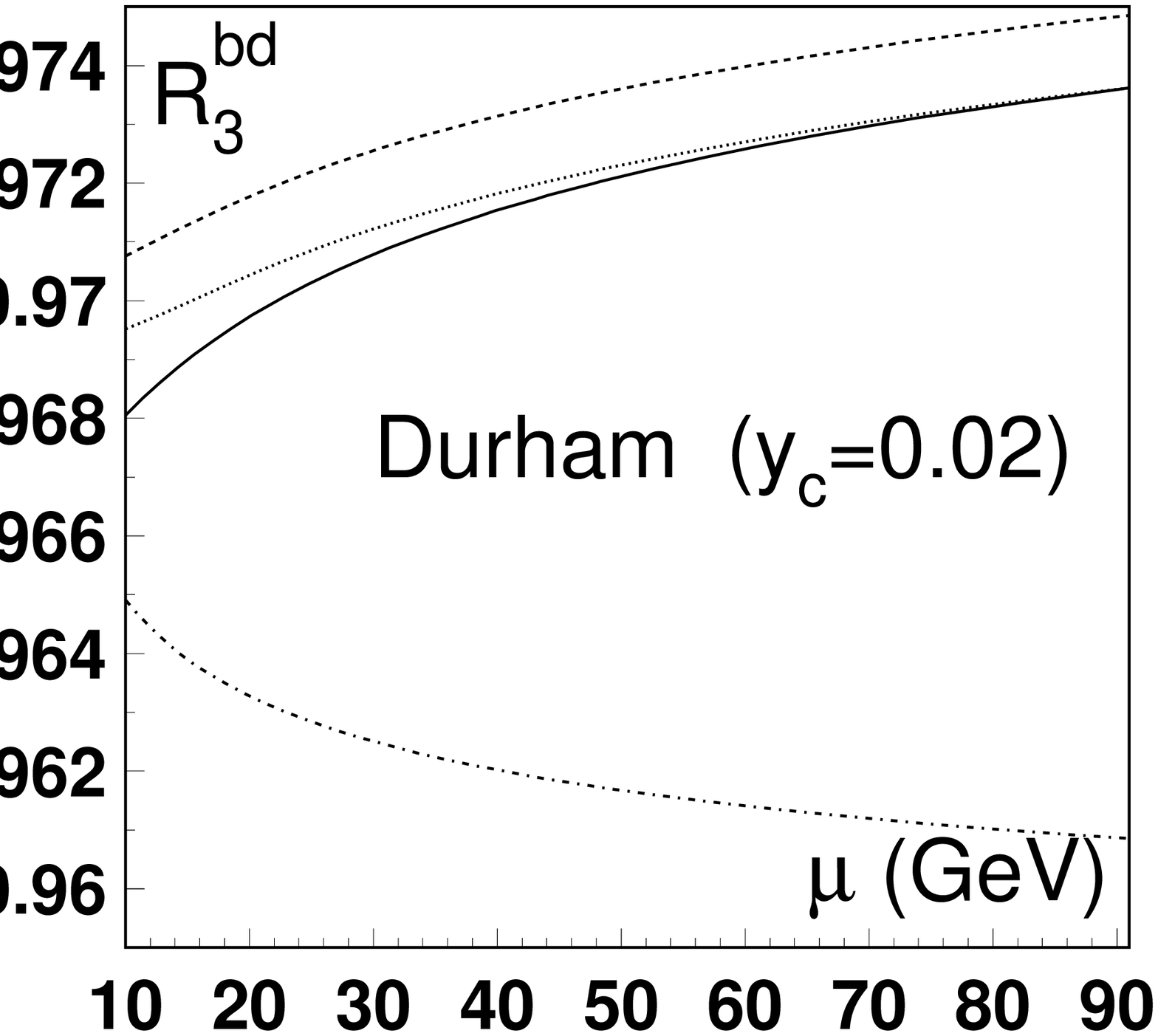}{-1.cm}
{\it The ratio $R_3^{bd}$
as a function of the scale $\mu$ for $y_c=0.02$.}{fig:r3bdmu}
%%%%%%%%%%%%%%%
\vskip -8mm
Other curves show $\mu$-dependence when
$R_3^{bd}$ is parameterized in terms of the running mass, $\overline{m}_b(M_Z)$,
eq.(\ref{r3bdrun}), but different mass definitions have been used in the logarithms.
The conservative estimate of the theoretical error for the $R_3^{bd}$
is to take the whole spread given by the curves.
The uncertainty in $R_3^{bd}$ induces an error in the measured mass
of the b-quark, $\Delta R_3^{bd}=0.004 \rightarrow \Delta m_b \simeq 0.23 GeV$.
This theoretical uncertainty is, however, below
current experimental errors, which are dominated by
fragmentation.

To conclude, the NLO calculation is necessary for accurate description of the
three-jet final state with massive quarks in $e^+e^-$-annihilation. 
Further studies of different observables and different jet-algorithms could
be very useful for the reduction the uncertainty of such calculation.

\noindent{\bf Acknowledgments.} 
We are indebted to S. Cabrera, 
J. Fuster and S. Mart\'{\i} for an 
enjoyable collaboration. M.B. is grateful to Laboratoire de Physique Math\'ematique
et Th\'eorique for the warm hospitality during his stay at Montpellier.

\end{document}